\documentclass{ifacconf}

\usepackage{graphicx}      
\usepackage{amsmath}       
\usepackage{amssymb}       
\usepackage{amsfonts}      
\usepackage{natbib}        
\usepackage{url}           
\usepackage{color}         
\usepackage{dsfont}        
\usepackage{comment}

\begin{document}
\begin{frontmatter}


\title{Image-Driven Control with Application to Thermoforming} 

\thanks{This work was supported by the Natural
Sciences and Engineering Research Council of Canada (NSERC) under
Grant RGPIN-2023-03660. 
© 2026 the authors. This work has been accepted to IFAC for publication under a Creative Commons Licence CC-BY-NC-ND.}
\author[First]{Junyong Shim}
\author[First]{Klaske van Heusden}

\address[First]{School of Engineering, The University of British Columbia, Kelowna, BC V1V 1V7, Canada (jun.shim@ubc.ca, klaske.vanheusden@ubc.ca)}

\begin{abstract}
This paper proposes Image-Driven Control (IDC), an approach to vision-based feedback that operates directly in the image space. IDC offers a direct end-to-end solution in the behavioral framework. Targeting a limited subset of vision-based feedback problems, including applications in manufacturing, the IDC design builds on a computationally advantageous subspace predictive control formulation and offset-free tracking to enable reference tracking in the image space under input and pixel-level constraints. IDC provides a control-theoretic alternative to vision-based feedback systems, enabling improved transparency for analysis and tuning of end-to-end control. It offers straightforward experiment design, an advantage in applications where data collection is expensive. The effectneriveness of IDC is shown in a simulation example using a nonlinear high-fidelity model of the heating phase of the thermoforming process.
\end{abstract}


\begin{keyword}
Vision-based feedback; Data-driven control theory;
Subspace predictive control; 
Thermal processes
\end{keyword}
\end{frontmatter}

\section{Introduction}
Machine vision allows for measurements, and consequently feedback control, of processes previously not accessible to conventional sensing. High-resolution optical and infrared cameras can monitor weld pools \citep{mobaraki2024vision}, melt pools in additive manufacturing \citep{khanafer2024condition}, and temperature fields in thermoforming (see, e.g., \cite{hosseinionari2023development,hosseinionari2025towards}). Vision-based feedback has been explored extensively in robotics and autonomous systems, from visual servoing \citep{chaumette2016visual} to end-to-end visuomotor policies \citep{finn2017deep,amini2020learning} and vision-language-action (VLA) models \citep{kim2024openvla}. However, its use remains limited in other applications such as manufacturing. 
 
Most vision pipelines are based on machine learning and feature extraction. Images are mapped to a small set of derived variables, which are then used for control. This pipeline requires data sets that are expensive to obtain in manufacturing applications. Feature extraction is usually designed or trained separately from the controller, without guarantees that the selected features are control-relevant. End-to-end deep learning solutions \citep{levine2016end,finn2017deep} combine perception and control, but rely on complex models with limited transparency for analysis and tuning. Trained on large task-specific data sets, these methods require substantial fine-tuning when the task changes \citep{chen2024end}, requiring additional expensive data collection and limiting their use in manufacturing applications. 

\cite{lu2021image} proposed a Dynamic Mode Decomposition based MPC framework (DMD-MPC) that builds a low-dimensional model from image sequences and embeds this in MPC. DMD-MPC assumes full state measurements in each image. The reduced-order design is tractable, but control design and analysis are performed on a compressed surrogate rather than in the full image space.

We propose \emph{Image-Driven Control} (IDC), a control-theoretic approach to vision-based feedback that operates directly in the image space. The proposed IDC in the behavioral framework targets vision-based feedback applications, including those with thermal imaging measurements, for which the sequence of past images contains sufficient control-relevant information. Unlike DMD-MPC, IDC does not assume that each image is a full state measurement; rather, the measured image trajectory is used as the feedback information. This requirement is not specific to IDC: any vision-based feedback system requires the measured image, or extracted features, to contain the information needed for control. The proposed controller is linear and therefore most appropriate when a local linear description is reasonable, for example operation around a fixed setpoint covered by the identification data. IDC allows for constraint handling and analysis directly in the image space, and for straightforward experiment design in manufacturing settings where experiments are expensive. The efficacy of IDC is illustrated in simulation using a high-fidelity nonlinear simulator of the heating phase of a thermoforming process, showing proof-of-principle.

The paper is organized as follows: Section~\ref{sec:prelim} describes data-driven predictive control and subspace predictive control; Section~\ref{sec:IDC} introduces IDC; Section~\ref{sec:simulation} presents the thermoforming simulation study, followed by the conclusion.

\section{PRELIMINARIES}
\label{sec:prelim}
Consider a discrete-time, linear time-invariant (DLTI), strictly proper system
\begin{equation}
\label{eq:ss_model}
x_{k+1} = A x_k + B u_k, \qquad
y_k = C x_k,
\end{equation}
with state $x_k \in \mathbb{R}^n$, input $u_k \in \mathbb{R}^m$, and output
$y_k \in \mathbb{R}^\mathcal{I}$.

\noindent\textit{Assumption~1 (Minimal realization).}
$(A,B)$ is controllable and $(C,A)$ is observable. Consequently, $n$ is the order of the system.

For a prediction horizon $N$, we define the stacked future outputs and inputs as
\begin{align*}
\bar{y} &= \operatorname{col}(y_{k+1\mid k},\ldots,y_{k+N\mid k})
   \in \mathbb{R}^{\mathcal{I}N},\\
\bar{u} &= \operatorname{col}(u_k,\ldots,u_{k+N-1})
   \in \mathbb{R}^{mN}.
\end{align*}
where $\operatorname{col}(\cdot)$ denotes vertical concatenation of its arguments, and $\bar{y}_{k+i|k}$ denotes the prediction of the output at time $k+i$ based on information available at time $k$.

The stacked prediction model can then be written as
\begin{equation}
\label{eq:model_mpc_prediction}
\bar{y} = \mathcal{O}_N x_k + \mathcal{T}_N \bar{u},
\end{equation}
where $\mathcal{O}_N$ is the $N$-step extended observability matrix and $\mathcal{T}_N$ is a lower block Toeplitz matrix constructed from $(A,B,C)$ (see, e.g., the batch formulation $\chi = S^x x_k + S^u U_0$ in \cite[Sec.~8.2]{Borrelli_Bemporad_Morari_2017}). In conventional model-based MPC, this prediction model is embedded in a quadratic program (QP) in $\bar{u}$ to track a reference trajectory subject to input and output constraints.

\subsection{Data-driven Predictive Control}
For an integer $L \ge 1$, the block Hankel matrix of a sequence $x_t^d \in \mathbb{R}^r$ is defined as
\begin{equation}
\label{eq:Hankel_def}
H_L(x) :=
\begin{bmatrix}
x_0^d     & x_1^d     & \cdots & x_{T-L}^d   \\
x_1^d     & x_2^d     & \cdots & x_{T-L+1}^d \\
\vdots    & \vdots    & \ddots & \vdots      \\
x_{L-1}^d & x_L^d     & \cdots & x_{T-1}^d
\end{bmatrix}
\in \mathbb{R}^{rL\times (T-L+1)}.
\end{equation}

Consider a noise-free input–output trajectory of \eqref{eq:ss_model}
\(
\{u_t^d,y_t^d\}_{t=0}^{T-1},
\)
where $u_t^d \in \mathbb{R}^m$ and $y_t^d \in \mathbb{R}^\mathcal{I}$. We write $H_L(u)$ and $H_L(y)$ for the input and output Hankel matrices.

\textit{Definition~1 (Persistence of excitation (PE) \citep{willems2005note}).}
An input sequence \(\{u_t^d\}_{t=0}^{T-1}\) is persistently exciting of order \(L\) if
\[
\operatorname{rank} H_L(u)=mL.
\]

\textit{Assumption~2 (Data and excitation).}
The data $\{u_t^d,y_t^d\}_{t=0}^{T-1}$ are generated by the DLTI system \eqref{eq:ss_model}, and the input sequence $u^d$ is persistently exciting of order $L + n$ and the data length $T$ is chosen large enough, i.e., $T \ge (m+1)(L+n)-1$.

Under \textit{Assumption 2}, the fundamental lemma \citep{willems2005note} states that a finite sequence $\{\Tilde{u}_k,\Tilde{y}_k\}_{k=0}^{L-1}$ is a trajectory of \eqref{eq:ss_model} if and only if there exists a coefficient vector $g \in \mathbb{R}^{T-L+1}$ such that
\begin{equation}
\label{eq:FL_compact}
\begin{bmatrix}
H_L(u) \\[0.2em] H_L(y)
\end{bmatrix} g
=
\begin{bmatrix}
\Tilde{u} \\[0.2em] \Tilde{y}
\end{bmatrix}.
\end{equation}

Building on this behavioral framework, Data-Enabled Predictive Control (DeePC) has been proposed as a trajectory-based predictive control method that constructs predictions directly from measured input–output data~\citep{8795639}.

Choose an initialization length $T_{\mathrm{ini}} \ge n$ and a prediction horizon $N$, and set $L := T_{\mathrm{ini}} + N$, for $L \le T$, such that the past input–output window uniquely determines the internal state. In DeePC, partition the Hankel matrices of input and output into past and future block rows as
\begin{equation}
\label{eq:UpYpUfYf_def}
\begin{bmatrix}
U_p \\[0.2em] U_f
\end{bmatrix}
:= H_L(u),
\qquad
\begin{bmatrix}
Y_p \\[0.2em] Y_f
\end{bmatrix}
:= H_L(y),
\end{equation}
where
\[
U_p \in \mathbb{R}^{mT_{\mathrm{ini}}\times (T-L+1)}, \quad
Y_p \in \mathbb{R}^{\mathcal{I}T_{\mathrm{ini}}\times (T-L+1)}
\]
collect the first $T_{\mathrm{ini}}$ block rows (past data), and
\[
U_f \in \mathbb{R}^{mN\times (T-L+1)}, \quad
Y_f \in \mathbb{R}^{\mathcal{I}N\times (T-L+1)}
\]
collect the remaining $N$ block rows (future data).

At time $k$, define the stacked past inputs and outputs as
\begin{equation}
\begin{aligned}\label{eq:uyini}
u_{\mathrm{ini}} &= \operatorname{col}(u_{k-T_{\mathrm{ini}}},\ldots,u_{k-1}),\\
y_{\mathrm{ini}} &= \operatorname{col}(y_{k-T_{\mathrm{ini}}},\ldots,y_{k-1}).
\end{aligned}
\end{equation}
and denote
\begin{equation}
\label{eq:uypredict}
\begin{aligned}
\bar{u} &= \operatorname{col}(u_k,\ldots,u_{k+N-1})\\
\bar{y} &= \operatorname{col}(y_{k\mid k},\ldots,y_{k+N-1\mid k})
\end{aligned}
\end{equation}
the future input and (predicted) output sequences. DeePC enforces the past-compatibility constraint
\begin{equation}
\label{eq:DeePC_const}
\begin{bmatrix}
U_p \\ Y_p
\end{bmatrix} g
=
\begin{bmatrix}
u_{\mathrm{ini}} \\ y_{\mathrm{ini}}
\end{bmatrix},
\end{equation}
and treating $g$ as the decision variable, it defines the future input and output trajectories as
\begin{equation}
\label{eq:DeePC_uy}
\bar{u} = U_f g, 
\qquad 
\bar{y} = Y_f g.
\end{equation}
The optimal trajectory is obtained by solving
\begin{equation}
\label{eq:DeePC_qp_g}
\begin{aligned}
\min_{g}\;\;
& \|y^{\mathrm{ref}}-\bar{y}\|_Q^2
  + \|\bar{u}\|_R^2 \\[0.2em]
\text{s.t.}\;\;
& \eqref{eq:DeePC_const},\eqref{eq:DeePC_uy},\quad
  \bar{u} \in \mathcal{U},\quad
  \bar{y} \in \mathcal{Y},
\end{aligned}
\end{equation}
where $\mathcal{U}$ and $\mathcal{Y}$ encode constraints. In the noise-free case, \eqref{eq:DeePC_qp_g} is equivalent to a finite-horizon MPC problem built on the prediction model \eqref{eq:model_mpc_prediction} \citep{9992453}.

\textit{Remark 1.} The data partition in \eqref{eq:UpYpUfYf_def}--\eqref{eq:uypredict} uses a past window ending at
time \(k-1\). If the current measurement \(y_k\)
is available when computing \(u_k\), a shifted partition can
instead include \(y_k\). This is an alternative lifted-state representation that conditions the predictor on the most recent output measurement. While \eqref{eq:UpYpUfYf_def}--\eqref{eq:uypredict} provide a sufficient statistic when the underlying linear representation is exact, using the most recent measurement may improve performance with finite data or model mismatch.

\subsection{Subspace Predictive Control (SPC)}
It has been shown that the modeling and control objectives can always be separated in the DeePC framework \citep{chiuso2025harnessing}. As the proposed IDC works with high-dimensional data, we adopt the 2-norm regularization that is equivalent to Subspace Predictive Control (SPC), offering computational advantages \citep{mattsson2023regularization}. In SPC, a linear regression model that maps past inputs/outputs and future inputs to future outputs is first identified and then used as a fixed predictor \citep{DINKLA2026112556,9992453}. Define the stacked regressor
\begin{equation}
\label{eq:W_z_def}
W :=
\begin{bmatrix}
U_p \\[0.2em] Y_p \\[0.2em] U_f
\end{bmatrix},
\qquad
z :=
\begin{bmatrix}
u_{\mathrm{ini}} \\[0.2em] y_{\mathrm{ini}} \\[0.2em] \bar{u}
\end{bmatrix}.
\end{equation}
Here $u_{\mathrm{ini}},y_{\mathrm{ini}},\bar{u}$ are defined as in \eqref{eq:uyini}--\eqref{eq:uypredict}. 

Given linearity in \eqref{eq:ss_model}, there exists a matrix $\Theta \in \mathbb{R}^{\mathcal{I}N\times(mT_{\mathrm{ini}}+\mathcal{I}T_{\mathrm{ini}}+mN)}$ such that, in the deterministic case,
\begin{equation}
\label{eq:regression_matrix_compact}
Y_f = \Theta\,W.
\end{equation}
In the presence of noise, $\Theta$ can be estimated by the least-squares problem
\begin{equation}
\label{eq:spc_ls_problem_compact}
\min_{\Theta} \; \big\| Y_f - \Theta\,W \big\|_2^2,
\end{equation}
which admits the solution
\begin{equation}
\label{eq:Theta_hat_compact}
\hat{\Theta} = Y_f\,W^\dagger,
\end{equation}
with $W^\dagger$ the Moore--Penrose pseudoinverse of $W$.

Partition
\begin{equation}
\label{eq:Theta_partition_compact}
\hat{\Theta}
=
\begin{bmatrix}
\hat{\Theta}_{\mathrm{ini}} & \hat{\Theta}_u
\end{bmatrix},
\end{equation}
with $\hat{\Theta}_{\mathrm{ini}} \in
\mathbb{R}^{\mathcal{I}N\times(mT_{\mathrm{ini}}+\mathcal{I}T_{\mathrm{ini}})}$ and $\hat{\Theta}_u \in \mathbb{R}^{\mathcal{I}N\times mN}$. The resulting predictor is 
\begin{equation}
\label{eq:spc_indirect_predictor_compact}
\bar{y}
=
\hat{\Theta}\,z
=
\hat{\Theta}_{\mathrm{ini}} z_{\mathrm{ini}}
+ \hat{\Theta}_u \bar{u},
\end{equation}
which has the same lifted structure as \eqref{eq:model_mpc_prediction}, but with an alternative state representation $z_{\mathrm{ini}} = \operatorname{col}(u_{\mathrm{ini}}, y_{\mathrm{ini}})$ identified directly from data instead of $x_k$. Under \textit{Assumption~2} and in the noise-free case, the predictions of \eqref{eq:spc_indirect_predictor_compact} coincide with those of DeePC \citep{fiedler2021relationship}. Note that in this formulation, the number of decision variables is reduced to $\bar{u}$.

\textit{Remark 2. (Offset-Free tracking)} Neither DeePC nor SPC, in their basic formulations, guarantees offset-free tracking. Classical offset-free MPC compensates such mismatch by augmenting the model with disturbance states or integrators. In the data-driven setting, offset-free behavior can be introduced through integrator-augmented formulation (using input increments as decision variables). The proposed IDC method builds on incremental SPC (iSPC) \citep{9992453}, but enforces constraints on the absolute inputs rather than on the increments.

\section{IMAGE-DRIVEN CONTROL METHODOLOGY}
\label{sec:IDC}
\subsection{Problem definition: Image feedback}
\label{sec:problemset}

We now specialize the DLTI setting of (1) to the case where
the measured output is an image rather than a small number
of output signals. We consider a plant with input $u_k \in \mathbb{R}^m$ and image output
\[
\mathbf{P} \in \mathbb{R}^{N_x \times N_y},
\]
where $N_x$ and $N_y$ denote the number of pixels (or grid cells) in the horizontal and vertical directions, respectively. The total number of pixels is $\mathcal{I} = N_x N_y,$
and \begin{equation}
\label{pixelsetup}
\mathbf{p} = \operatorname{vec}(\mathbf{P}) \in \mathbb{R}^\mathcal{I}
\end{equation}
is the stacked pixel vector used for control. In the DLTI case, this fits the structure
\begin{equation}
\label{eq:ss_pixels}
x_{k+1} = A x_k + B u_k, \qquad
\mathbf{p}_k = C x_k,
\end{equation}
with $x_k \in \mathbb{R}^n$ the (unmeasured) state and $\mathbf{p}_k$ the high-dimensional measured pixels (output image).

\noindent\textit{Assumption~3 (Pixel-level sensing).}
The plant admits a finite-dimensional minimal realization \eqref{eq:ss_pixels}, so that $(A,B)$ is controllable and $(C,A)$ is observable. Because $\mathbf{p}$ is typically large, we do \emph{not} assume output controllability with respect to all pixel values.

Assumption 3 states that the measured image trajectory contains sufficient information for feedback control. This is the image-space analogue of the information requirement in any vision-based feedback system: if the image does not contain the relevant system information,
neither direct image feedback nor feature extraction can recover the system state for control. The behavioral/SPC predictor is exact under the nominal
DLTI assumptions. For nonlinear plants, IDC should be interpreted as a local linear data-driven controller, suitable when closed-loop operation remains in a region where such an approximation is adequate.

\noindent\textbf{Image-Driven Control (IDC)} receives the complete image $\mathbf{P}$ (equivalently, $\mathbf{p}_k$) as feedback and aims to track a desired sequence of reference images while respecting bounds on the actuator inputs and admitting constraints on individual image pixels.

To explicitly condition the feedback law on the most recent image \(\mathbf{p}_k\) at control instant $k$, IDC uses the following shifted past/future partitioning:
\begin{equation}
\label{eq:IDC_windows}
\begin{aligned}
u_{\mathrm{ini}} &= \operatorname{col}(u_{k-T_{\mathrm{ini}}+1},\ldots,u_{k-1}),\\
\mathbf{p}_{\mathrm{ini}} &= \operatorname{col}(\mathbf{p}_{k-T_{\mathrm{ini}}+1},\ldots,\mathbf{p}_{k-1},\mathbf{p}_{k}),\\
\bar{u} &= \operatorname{col}(u_k,\ldots,u_{k+N}).
\end{aligned}
\end{equation}

Here, the boundary between past and future is shifted by one
step: the most recent output $\mathbf{p}_k$ is treated as known data,
and the first decision variable is $u_k$. This convention
conditions the feedback law on the most recent image measurement
while preserving the behavioral/SPC structure. Note that this
implementation differs from \citep{9992453} and does
not increase the number of past input samples used in the predictor.

Let $\{u_t,\mathbf{p}_t\}_{t=0}^{T-1}$ denote an open-loop data record and construct the following block Hankel matrices
\begin{equation}
\label{eq:UpYpUfYf_shift}
\begin{bmatrix}
U_p \\[0.2em] U_f
\end{bmatrix}
:= H_L(u),
\qquad
\begin{bmatrix}
P_p \\[0.2em] P_f
\end{bmatrix}
:= H_L(\mathbf{p}),
\end{equation}
with the shifted time window,
\[
U_p \in \mathbb{R}^{m(T_{\mathrm{ini}}-1)\times (T-L+1)}, \quad
P_p \in \mathbb{R}^{\mathcal{I}T_{\mathrm{ini}}\times (T-L+1)},
\]
and
\[
U_f \in \mathbb{R}^{m(N+1)\times (T-L+1)}, \quad
P_f \in \mathbb{R}^{\mathcal{I}N\times (T-L+1)}.
\]

\noindent\textbf{Offset-free IDC.}
Let the stacked future input increments over the horizon be
\begin{equation}
\Delta \bar{u} = \operatorname{col}(\Delta u_k,\ldots,\Delta u_{k+N}),
\end{equation}
where $\Delta u_{k+i} = u_{k+i} - u_{k+i-1},$

so that, for each $i = 0,\ldots,N$, $u_{k+i}=u_{k-1}+\sum_{j=0}^{i} \Delta u_{k+j}$.

Let $\bar{u}$ denote the stacked absolute inputs following the horizon in \eqref{eq:IDC_windows}. The relationship between $\bar{u}$ and $\Delta \bar{u}$ can be written compactly as
\begin{equation}
\label{eq:IA_sum_mapping}
\bar{u} = S_{\!\Sigma}\,\Delta \bar{u} + E\,u_{k-1},
\end{equation}
where
\[
S_{\!\Sigma} =
\begin{bmatrix}
I_m & 0   & \cdots & 0 \\
I_m & I_m & \ddots & \vdots \\
\vdots & \vdots & \ddots & 0 \\
I_m & I_m & \cdots & I_m
\end{bmatrix}\qquad
E =
\begin{bmatrix}
I_m \\ I_m \\ \vdots \\ I_m
\end{bmatrix}.
\]
Here $S\in \mathbb{R}^{m(N+1)\times m(N+1)}$, $E\in \mathbb{R}^{m(N+1)\times m}$, and $u_{k-1}$ is the most recent applied input, which is already contained in the last block of $u_{\mathrm{ini}}$ in $z_{\mathrm{ini}}$. Consequently, the term $E\,u_{k-1}$ can be absorbed into the contribution of $z_{\mathrm{ini}}$ in the predictor, without introducing an additional state.

\noindent\textbf{Prediction with input increments.}
With the shifted timing convention, the Hankel matrices
\eqref{eq:UpYpUfYf_shift} are used in an SPC-style regression
(cf.\ \eqref{eq:regression_matrix_compact}--\eqref{eq:spc_ls_problem_compact})
to obtain an estimate of future images \(P_f = \Theta W\).
Since the full image output may introduce redundant or linearly dependent
pixel information, and output controllability with respect to all pixel values
is not assumed, the corresponding pseudoinverse may be ill-conditioned and
is computed using truncated SVD in practice. Partitioning \(\Theta\) as in
\eqref{eq:Theta_partition_compact}, but with the shifted window, yields
\[
\Theta_{\mathrm{ini}} \in \mathbb{R}^{\mathcal{I}N\times\big(m(T_{\mathrm{ini}}-1)+\mathcal{I}T_{\mathrm{ini}}\big)}, 
\qquad
\Theta_u \in \mathbb{R}^{\mathcal{I}N\times m(N+1)}.
\]

Let
\[
\bar{\mathbf{p}}
:= \operatorname{col}(p_{k+1\mid k},\ldots,p_{k+N\mid k})
\in \mathbb{R}^{\mathcal{I}N}
\]
denote the stacked predicted pixel vector over the horizon. With
\(
z_{\mathrm{ini}} = \operatorname{col}(u_{\mathrm{ini}},\mathbf{p}_{\mathrm{ini}})
\)
and absolute future inputs $\bar{u}$, substituting the integration mapping \eqref{eq:IA_sum_mapping} gives
\begin{align}
\bar{\mathbf{p}}
&= \Theta_{\mathrm{ini}} z_{\mathrm{ini}}
 + \Theta_u\big(S_{\!\Sigma}\,\Delta \bar{u} + E\,u_{k-1}\big) \nonumber\\[0.2em]
&= \underbrace{\big(\Theta_{\mathrm{ini}} + \Theta_uE_{\mathrm{fold}}\big)}_{A_{\mathrm{ini}}}
   z_{\mathrm{ini}}
 + \underbrace{\Theta_u S_{\!\Sigma}}_{H}\,\Delta \bar{u},
\label{eq:IA_SPC_pred_folded}
\end{align}
where $E_{\mathrm{fold}}$ denotes the matrix (of the same size as $\Theta_{\mathrm{ini}}$) that maps $\Theta_u E$ into the columns of $\Theta_{\mathrm{ini}}$ corresponding to the last input block $u_{k-1}$.
In practice, $E_{\mathrm{fold}}$ does not need to be formed explicitly; its effect is implemented by adding $\Theta_u E$ to those columns. The matrix
\[
A_{\mathrm{ini}} \in \mathbb{R}^{\mathcal{I}N\times\big(m(T_{\mathrm{ini}}-1)+\mathcal{I}T_{\mathrm{ini}}\big)}
\]
thus captures both the original dependence on $z_{\mathrm{ini}}$ and the integrated contribution of $u_{k-1}$, while
\[
H = \Theta_u S_{\!\Sigma} \in \mathbb{R}^{\mathcal{I}N\times m(N+1)}
\]
maps the stacked input increments $\Delta \bar{u}$ to the predicted future images $\bar{\mathbf{p}}$.
Equation \eqref{eq:IA_SPC_pred_folded} has the same structural form in its prediction as \eqref{eq:spc_indirect_predictor_compact}, but now the integrator is embedded.

\vspace{0.3em}
\noindent\textbf{Image–Driven Control (IDC).}
The IDC optimization problem is defined as
\begin{equation}
\label{eq:IDC_QP}
\begin{aligned}
\min_{\Delta \bar{u}}\;\;
& \big\| \bar{\mathbf{p}}_{\mathrm{ref}} - \bar{\mathbf{p}} \big\|_{Q}^2
  + \|\Delta \bar{u}\|_{R_{\Delta}}^2 \\[0.3em]
\text{s.t.}\;\;
& \bar{u} = S_{\!\Sigma}\,\Delta \bar{u} + E\,u_{k-1}, \\[0.2em]
& \bar{\mathbf{p}} = A_{\mathrm{ini}} z_{\mathrm{ini}} + H\,\Delta \bar{u}, \\[0.2em]
& z_{\mathrm{ini}} = \operatorname{col}(u_{\mathrm{ini}}, \mathbf{p}_{\mathrm{ini}}), \\[0.2em]
& \bar{u} \in \mathcal{U},\;\; \bar{\mathbf{p}} \in \mathcal{P},
\end{aligned}
\end{equation}
where \(\bar{\mathbf{p}}_{\mathrm{ref}}\in\mathbb{R}^{\mathcal{I}N}\)
is the stacked reference image trajectory, ordered and vectorized
consistently with \(\bar{\mathbf{p}}\), and \(Q\succeq 0\) and
\(R_{\Delta}\succ 0\) are weighting matrices. The sets
\[
\mathcal{U} := [u_{\min},u_{\max}] \subset \mathbb{R}^{m(N+1)},\quad
\mathcal{P} := [\mathbf{p}_{\min},\mathbf{p}_{\max}]
\subset \mathbb{R}^{\mathcal{I}N}
\]
denote actuator and pixel box constraints. This formulation introduces explicit integral behaviour while preserving absolute input constraints. A key feature of the IDC formulation is that the decision
variable is the input-increment sequence \(\Delta \bar{u}\),
with dimension \(m(N+1)\), independent of the number of
pixels. The prediction matrices are computed offline. Thus, the online QP dimension scales with the input sequence, not the number of pixels, except for the number of constraints.

\section{SIMULATION example: heating phase of the thermoforming process}
\label{sec:simulation}

\subsection{System Description}
To illustrate the proposed methodology, we apply IDC to the heating phase of a thermoforming process. Thermoforming of thermoplastic sheets is widely used in automotive, aerospace, and packaging. Energy use, part quality, and consequently waste product depend on how well the sheet temperature field matches the desired mould-dependent profile \citep{hosseinionari2023development}. This process configuration is inherently multivariable and high-dimensional where spatial temperature information is highly valuable. Feasibility of feedback using linear MPC based on zone-averaged temperatures has been shown in a research experimental setup \citep{hosseinionari2025towards}. This supports thermoforming as a proof-of-principle case where local linear control can provide adequate performance for the nonlinear plant. Full-image feedback can enable region-dependent constraints, such as tighter bounds in critical forming zones, and may improve achievable spatial temperature distributions, supporting product quality and waste reduction. We adopt the high-fidelity nonlinear simulation model developed in \cite{hosseinionari2023development} as our numerical testbed, where the heat transfer involves radiative heating, in-plane conduction, and convective losses, modeled by a standard energy balance for each sheet element. The sheet surface is discretized on an $N_x \times N_y$ grid, where in the simulations $\mathcal{I} = N_x N_y = 23 \times 37 = 735,$ and an ideal IR camera is assumed to measure the corresponding temperature field, providing the output image $\mathbf{P} \in \mathbb{R}^{N_x \times N_y}$ used for feedback.

\subsection{Data collection and IDC design}
\label{subsec:implementation_details}

The IDC predictor is identified from a dedicated open-loop simulation of the high-fidelity nonlinear model with sampling period \(T_s = 60\,\mathrm{s}\). The system has $m = 15$ inputs and images contain $735$ pixels ($\mathcal{I} = 735$ outputs). An input trajectory of length $T_{\mathrm{id}}=10000$ is applied using a multi-input pseudo-random binary sequence (PRBS) with amplitude range $[0,200]$~W, starting from a uniform $25^\circ$C sheet, and the simulator returns the temperature image $\mathbf{P}_k$ at each sampling instant. The resulting dataset $\{(u_k,\mathbf{p}_k)\}_{k=0}^{T_{\mathrm{id}}-1}$ is stored and used to define the image-driven controller. For the corresponding DLTI setting, \(T_{\rm id}\) exceeds the
lower bound associated with the PE condition in Assumption 2.
Since the simulator is nonlinear, this condition is not interpreted
as a guarantee of exact predictive validity; rather, the data are
used to construct a finite-data local linear predictor over the
operating region explored in identification.

Prior to forming Hankel matrices, the steady-state resulting from the environment of $25^\circ$C is subtracted from each image. The trajectory is then arranged into Hankel matrices \eqref{eq:UpYpUfYf_shift} with past window $T_{\mathrm{ini}} = 10$ and prediction horizon $N_p = 20$, using the shifted construction consistent with \eqref{eq:IDC_windows}. A linear image-driven predictor is obtained by solving a least-squares problem \eqref{eq:regression_matrix_compact}--\eqref{eq:spc_ls_problem_compact} based on the stacked regressor $W := \begin{bmatrix} U_p^\top & P_p^\top & U_f^\top \end{bmatrix}^\top$, implemented via a truncated singular value decomposition of \(W\). Writing \(W=USV^\top\), with singular values \(\{\sigma_i\}\)
on the diagonal of \(S\), we retain those satisfying
\(\sigma_i \geq 10^{-5}\sigma_{\max}\), corresponding to 677
singular values, and form the pseudoinverse using only the
retained singular values. We designed the IDC with $Q = 10 I_{\mathcal{I}N_p}$ and $R_{\Delta} = 0.5 I_{m(N_p+1)}$. The decision variable is $\Delta \bar{u} \in \mathbb{R}^{m(N_p+1)=315}$. At each sampling instant, this QP is solved with input constraints $[0,500]$W.

\begin{figure*}[t]
  \centering
  \includegraphics[width=0.9\textwidth]{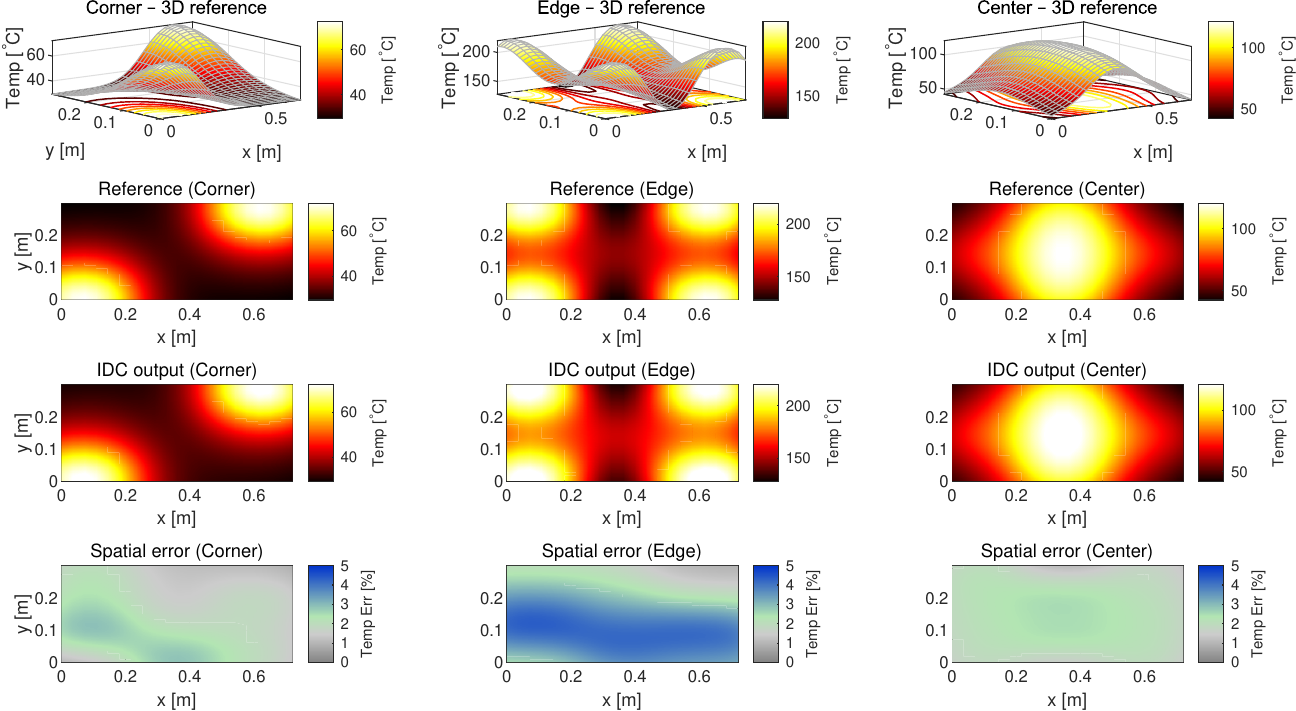}
    \caption{Steady-state reference temperature fields and IDC closed-loop outputs for three heating patterns (corner, edge, center): 3D surfaces (top), 2D projections (second row), steady-state IDC outputs (third row), and spatial steady-state error on a unified percentage scale (bottom).}
  \label{fig:IDC_ref_heatmaps}
\end{figure*}

\begin{figure*}[t]
  \centering
  \includegraphics[width=0.9\textwidth]{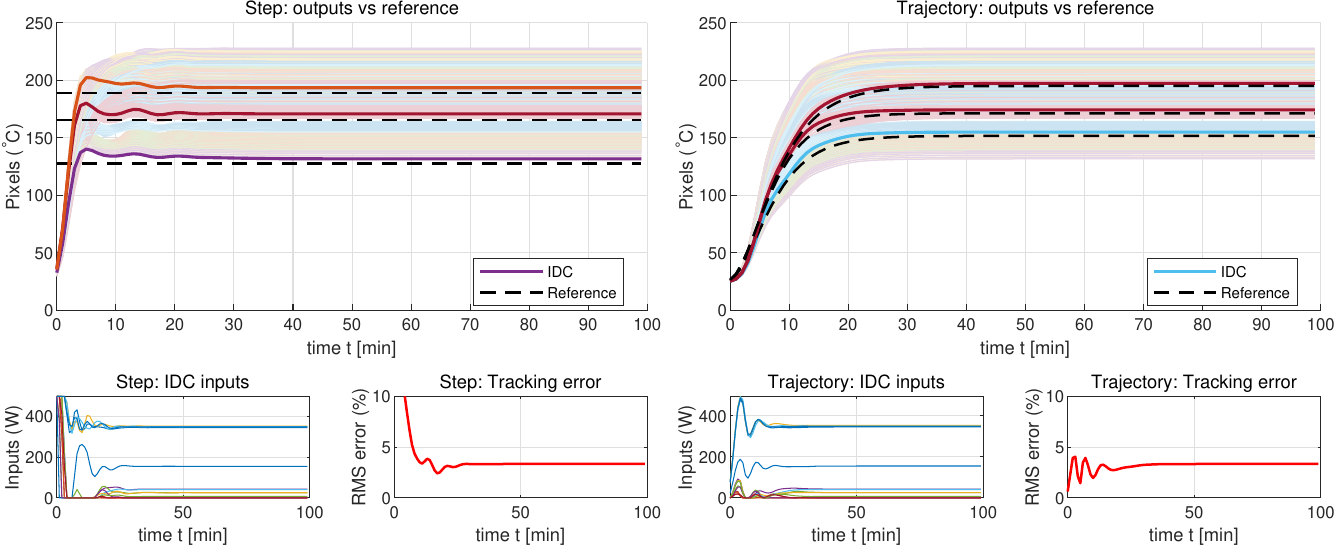}
  \caption{Closed-loop IDC performance for the edge heating pattern under step (left) and trajectory (right) references: all pixel temperatures and three representative trajectories (top), heater inputs and RMS tracking error (bottom).}
  \label{fig:IDC_flagship}
\end{figure*}

\subsection{Closed-loop IDC performance}
Figure~\ref{fig:IDC_ref_heatmaps} shows three steady-state reference temperature fields used as targets in the simulations. Each pattern represents a distinct spatial heating objective on the sheet, ranging from a localized hot corner to an edge-focused profile and a centrally concentrated field. All reference fields are obtained by simulating the nonlinear heater–sheet model. Thus, each reference is known to be reachable for the testbed model. The subplots below in Fig.~\ref{fig:IDC_ref_heatmaps} show the final IDC output images for each pattern. The bottom plots display the spatial tracking error on a unified percentage colour scale.

Figure~\ref{fig:IDC_flagship} shows the closed-loop IDC behaviour for the ``edge'' pattern under step and trajectory references. Three representative pixels (from low, medium, and high reference levels) are highlighted together with their references. The trajectories converge around their respective reference values. The bottom panels show heater inputs and RMS tracking error plotted on a common scale. Inputs stay within bounds and settle to steady levels, while the error decays and remains small. These results demonstrate the feasibility
of applying IDC as a local linear predictive controller on this nonlinear
simulation testbed.

The closed-loop trajectories exhibit a small but non-zero steady-state offset. In the present study, the IDC predictor is a linear model identified from data, whereas the underlying heater–sheet dynamics are highly nonlinear and coupled. This inherent model–plant mismatch means that the offset-free property of the integrator-augmented formulation holds only approximately; errors in the prediction manifest as residual tracking error in the image domain. The closed-loop performance depends on several identification and discretization choices made in the setup, including the PRBS input design, the Hankel window lengths, the SVD truncation level, and the controller sampling time (cf. Sec.~\ref{subsec:implementation_details}), highlighting the importance of experiment design and tuning for application of IDC. The present study demonstrates feasibility of IDC and robustness to a plant-model mismatch, assuming ideal sensing. Robustness to noise and disturbances remains an important direction for future work.
\section{CONCLUSION}
This paper presents a proof-of-principle implementation of image-driven control (IDC), demonstrating that IDC is both effective and computationally tractable. The results show that full-field images can be used directly in a predictive control framework, without learning-based feature extraction. Since IDC is formulated entirely within a data-driven control methodology, analysis tools for control-relevant properties such as controllable output subspaces can be applied directly in the image space. As the considered vision-based feedback framework is generally not output controllable, future work includes the design of high-dimensional objectives and analysis of control performance with respect to such objectives. Further directions include experimental validation, robustness with respect to noise and experiment design.


\bibliography{ifacconf}

\end{document}